\documentclass{webofc}

\usepackage[varg]{txfonts}   
\usepackage{hyperref}
\usepackage{url}
\hypersetup{colorlinks=true,citecolor=blue,urlcolor=blue,linkcolor=blue}
\begin{document}
\title{Recent results from NA61/SHINE}
%
%

\author{\firstname{Andrzej} \lastname{Rybicki}\inst{1}\fnsep\thanks{\email{andrzej.rybicki@ifj.edu.pl}} 
       \lastname{for the NA61/SHINE Collaboration}
}

\institute{{Institute of Nuclear Physics, Polish Academy of Sciences, Radzikowskiego 152, 31-342 Krak\'ow, Poland}
          }

\abstract{The NA61/SHINE experiment at the CERN SPS is a multipurpose fixed-target spectrometer for charged and neutral hadron measurements. Its research program includes studies of strong interactions as well as reference mea\-surements for neutrino and cosmic-ray physics. A significant advantage of NA61/SHINE over collider experiments is its extended coverage of phase space available for hadron production. This includes the nearly entire forward hemi\-sphere for charged hadrons and additionally, a large part of the backward hemisphere for specific neutrals. This paper summarizes a selected set of new results, obtained by NA61/SHINE since the last SQM conference (Busan, 2022). Particular attention is devoted
  to (1) the first-ever direct measurement~of open charm production in nucleus-nucleus collisions at SPS energies (2) the difference observed between charged and neutral meson production in Ar+Sc reactions, up to now not understood by existing models, and (3) the importance of baseline effects in the search for the critical point of strongly interacting matter.}
\maketitle
      \vspace*{-0.6mm}
\section{Introduction}
\label{intro}
NA61/SHINE~\cite{1} is the second largest fixed target experiment at CERN. As a direct, upgraded continuation of its predecessor NA49, the detector offers a nearly continuous experimental data flow for three decades. Its research program includes heavy ion physics as well as measurements for neutrino (J-PARC, Fermilab) and cosmic-ray studies (Pierre-Auger, KASCADE, satellite experiments). In this paper, only a selected set of most recent results
related to the NA61/SHINE strong interaction program will be addressed.
      \vspace*{-0.1mm}
\section{Open charm physics at the SPS}
\label{charm}
\begin{figure}[h]
\centering
\sidecaption
\includegraphics[width=7cm,clip]{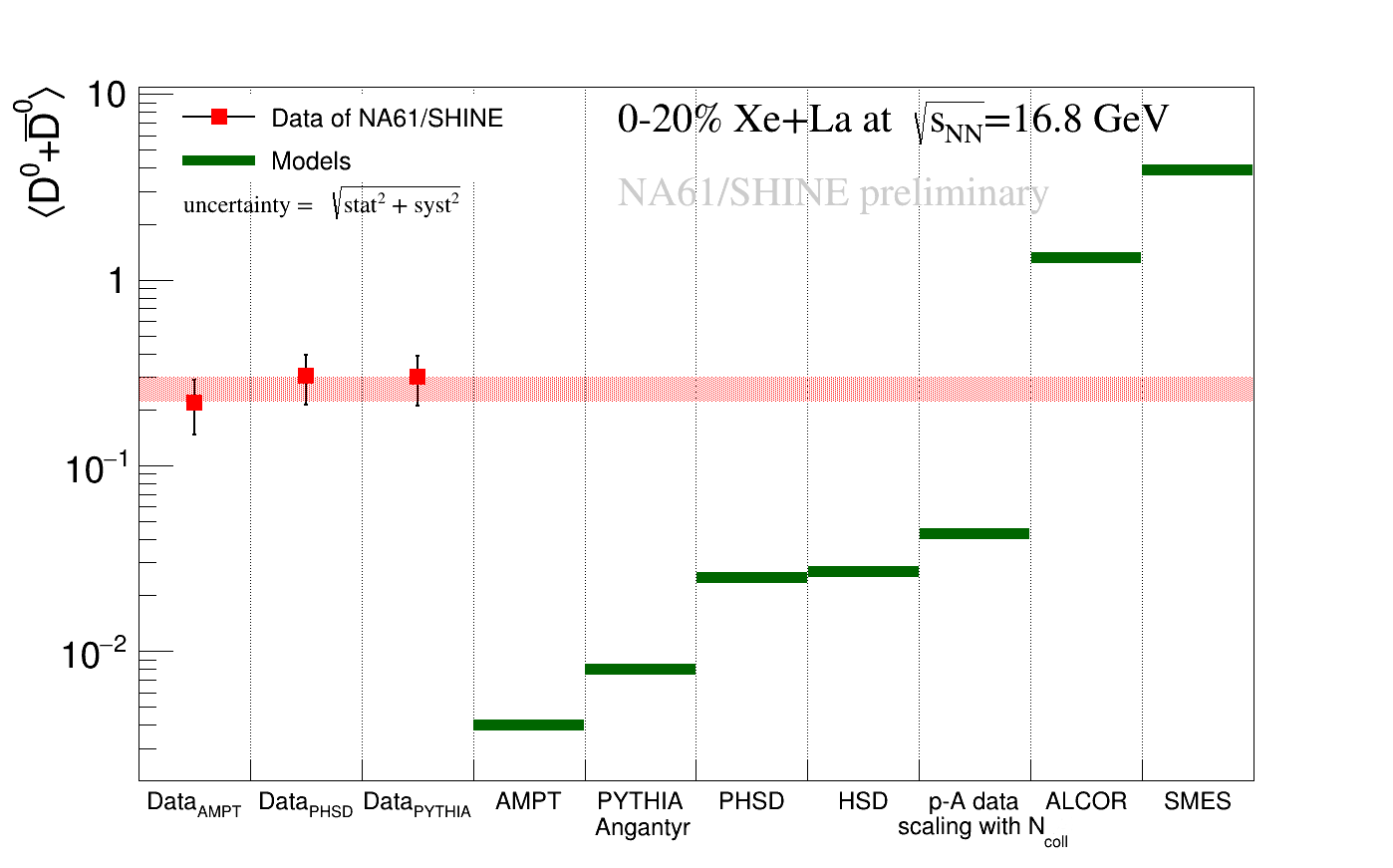}
\caption{The open charm 4$\pi$ yield $\langle D^0+\overline{D}{\hspace{0.1mm}}^0\rangle$ in central Xe+La collisions at $\sqrt{s_{NN}}$=16.8~GeV, put in comparison to predictions of theoretical models. The red band gives the theoretical (model-related) uncertainty. The figure comes from Ref.~\cite{a} (see therein for more details).}
\label{f2}       
\end{figure}

The first-ever direct measurement of open charm production in nucleus-nucleus collisions at CERN SPS energies has recently been performed by NA61/SHINE. This result, obtained in Xe+La collisions at $\sqrt{{s}_{NN}}$=16.8~GeV, is described in detail in Ref.~\cite{a}. The $D^0+\overline{D}{\hspace{0.1mm}}^{0}$ yield visible in acceptance was extrapolated to the total 4$\pi$ yield using three different models to estimate the corresponding model-related uncertainty. As apparent in Fig.~\ref{f2}, the NA61/SHINE measurement offers sufficient discrimination power to bring a decisive improvement to existing theoretical models of open charm production, which in their present version differ by as much as three orders of magnitude at this collision energy.

The second direct measurement in A+A reactions, and the first-ever for open charm in heavy ion collisions at the SPS, is also in progress in the framework of the NA61/SHINE open charm program with the upgraded detector. The corresponding data-taking of Pb+Pb collisions started in 2022 and
is planned to
be completed in 2025. The accumulated statistics will allow for the first measurements of open charm as a function of Pb+Pb collision centrality, rapidity, or transverse momentum ever made at SPS energies. This will further significantly improve the theoretical knowledge of charm production mechanisms close to threshold. A continuation of charm studies at the SPS with still higher statistics was also proposed by the NA60+ Collaboration, in a Letter of Intent at the end of 2022~\cite{2}.

\begin{figure*}[b!]
\centering
\vspace*{0cm}       
\hspace*{-0.6cm}
\includegraphics[width=6.05cm,clip]{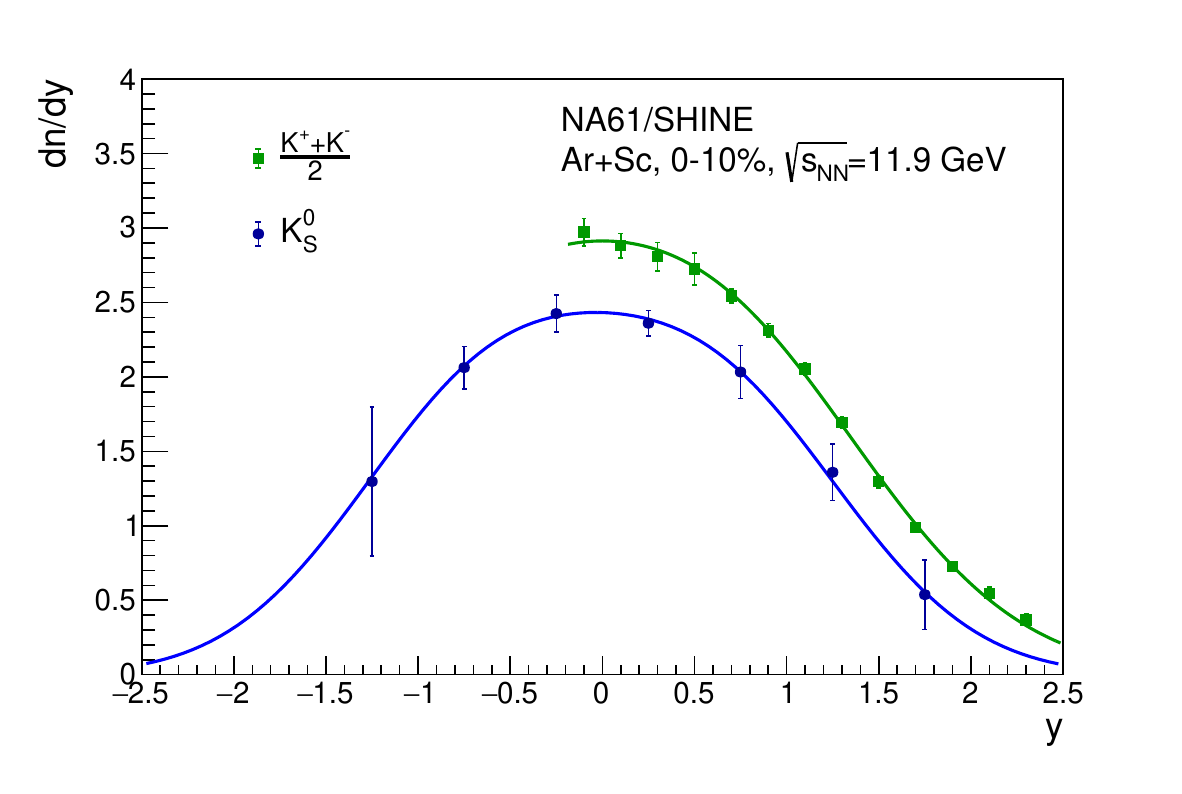}
\vspace*{-0.2cm}       
\hspace*{-0.44cm}
\includegraphics[width=7.48cm,clip]{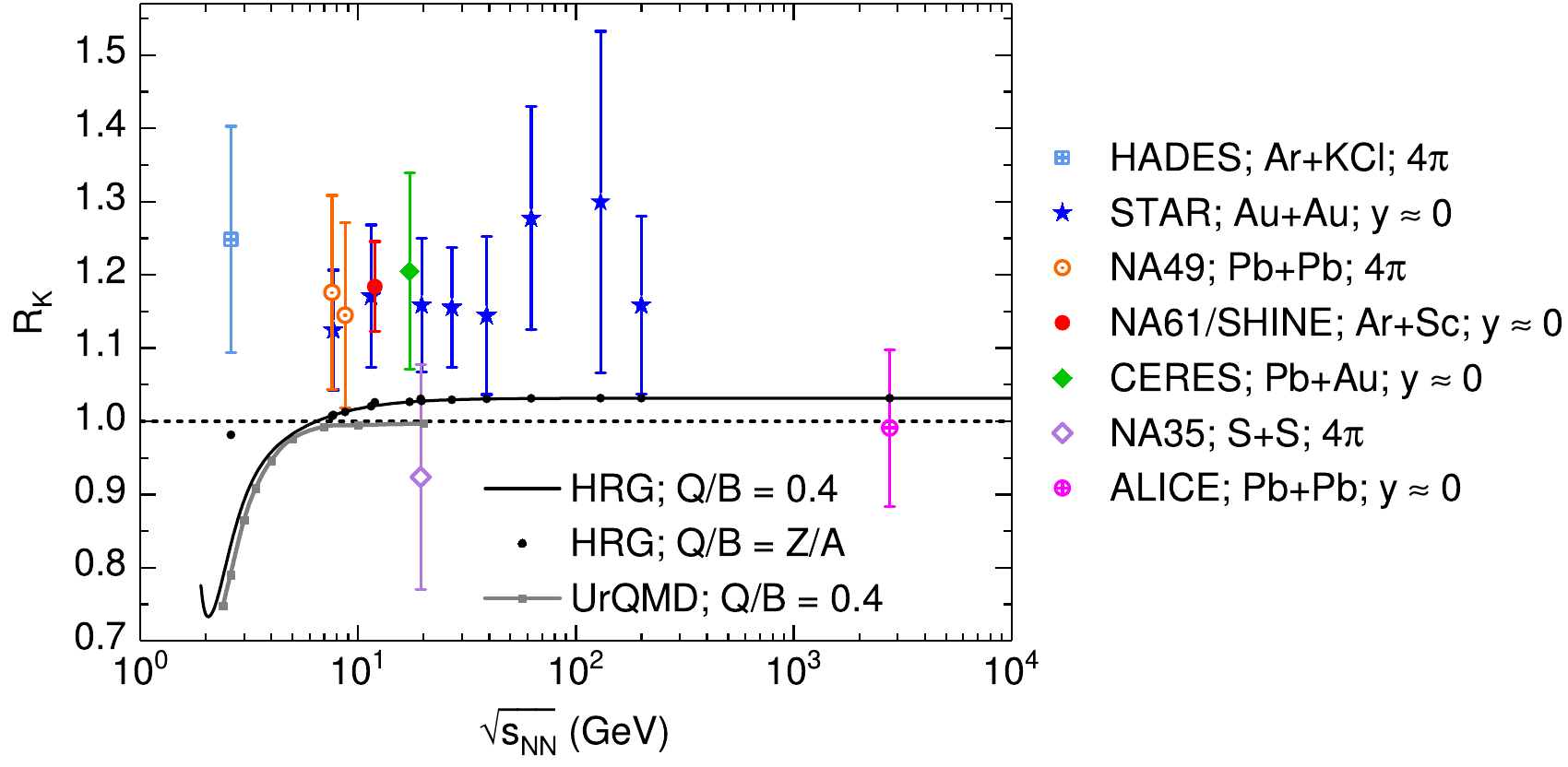}
\hspace*{-0.6cm}
  \begin{picture}(1,1)
 \put(-283,25){(a)}
 \put(-98,81){(b)}
  \end{picture}
  \caption{(a) Comparison of rapidity spectrum of neutral kaons ($K^0_S$) with the averaged spectrum of charged kaons ($K^++K^-$)/2 in Ar+Sc collisions. From Ref.~\cite{3}. (b) Ratio $R_K\,$=$\,(K^++K^-)$/(2$K^0_S$), drawn as a function of$\sqrt{s_{NN}}$ for NA61/SHINE data and
    the
    compilation of results from earlier experiments. The compilation includes ratios of mid-rapidity densities
as well as
4$\pi$ yields, see \mbox{Ref.~\cite{3} for a full biblio}\-graphy.
  {Experimental data is compared to predictions of UrQMD and HRG models ($Q/B$ is the assumed} {charge-to-baryon number ratio). From Ref.~\cite{4}, updated. In both plots total uncertainties are drawn.}}
\label{f3}       
\end{figure*}
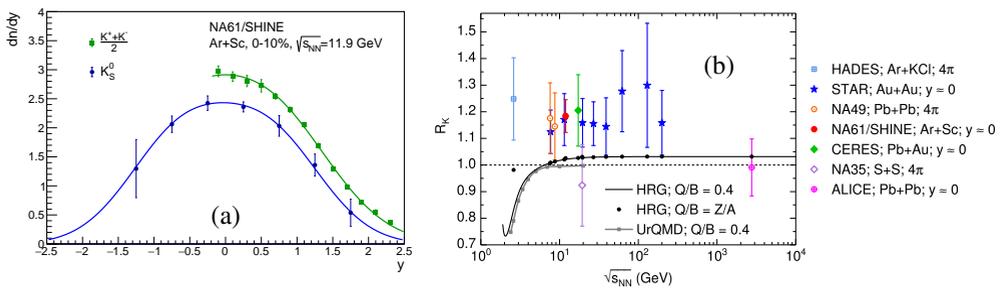

\section{Unexpected excess of charged over neutral kaons}
\label{iso}
A simultaneous, large-acceptance measurement of $K^+$, $K^-$, and $K^0_S$ production in Ar+Sc collisions at $\sqrt{s_{NN}}$=11.9~GeV has recently been performed~\cite{3a,3} (see also a detailed account in Ref.~\cite{5}). A comparison of averaged charged to neutral kaon rapidity spectra
is shown in Fig.~\ref{f3}~(a). A clear excess of charged kaon production is apparent. Isospin symmetry between $u$ and $d$ quarks, which can be regarded as a specific case of the flavour symmetry of QCD in the massless quark limit, predicts the equality $(K^++K^-)/2=(K^0+\overline{K}{\hspace{0.1mm}}^{0})/2=K^0_S$ for „charge-symmetric” nuclei ($Z=N\equiv A-Z$) once we neglect the very small effect of CP violation. The observed inequality of charged and neutral kaon yields does not follow this prediction. As such, it is to be concluded that this constitutes an evidence of an unexpectedly large violation of isospin symmetry, or flavour symmetry, in the high-energy Ar+Sc reaction.\footnote{The Ar and Sc nuclei are not exactly charge-symmetric. The moderate excess of neutrons over protons is predicted to decrease charged kaon compared to neutral kaon production, contrary to the result shown in Fig.~\ref{f3}~(a).}

A comparison of the NA61/SHINE result to a compilation performed on the basis of other existing data is shown in Fig.~\ref{f3}~(b). Although with large uncertainties for the individual measurements, the latter consistently support the Ar+Sc result in the range {$2.6\,$$<$\hspace*{-0.5mm}$\sqrt{{s}_{NN}}$\,$<$\,$200$~GeV}. The observed effect cannot be understood on the basis of UrQMD and HRG models, including known effects of isospin symmetry violation (see also Ref.~\cite{4}).

An unexpected effect of strong violation of isospin symmetry in multi-particle production can be of fundamental importance for the understanding of the strong force. This calls for a concentrated experimental and theoretical effort.

On the experimental side, it is interesting to note that the experimental uncertainties of the earlier measurements, Fig.~\ref{f3}\,(b) appear larger than those of NA61/SHINE. This is in spite of the larger statistics of data published by collider experiments. Evidently, this question requires clarification from the respective Collaborations (the discussion of NA61/SHINE uncertainties can be found in Refs.~\cite{3a,3}). We note that a part of this problem could originate from the fact that collider measurements of $\mathrm{d}n/\mathrm{d}y$ yields at mid-rapidity can never be performed in the complete kinematic range due to their inherent $p_T-$cut off, which will increase their uncertainty. The ability to perform measurements starting at $p_T=0$ (sharp) is a clear advantage of NA61/SHINE.

Finally, we note that the excess of charged over neutral kaons has also been observed by NA61/SHINE in $\pi^-$+C
collisions.
This effect cannot be explained by microscopic \mbox{models}~(for more details, see Ref.~\cite{8}).
%
Therefore, in October 2024
the experiment
will perform a simultaneous measurement of a charge-symmetric ensemble of $\pi^+$+C and $\pi^-$+C reactions~\cite{9}.
This 
will allow to clarify whether the unexpected violation of flavour symmetry is unique to nucleus-nucleus collisions or is a more general property of strong interactions.

\section{Energy dependence of strangeness production}
\label{ene}

\begin{figure*}[h!]
\vspace*{-0.3cm}       
\sidecaption
  \centering
\vspace*{-0.9cm}       
\hspace*{-0.4cm}
\includegraphics[width=5.08cm,height=4.48cm,clip]{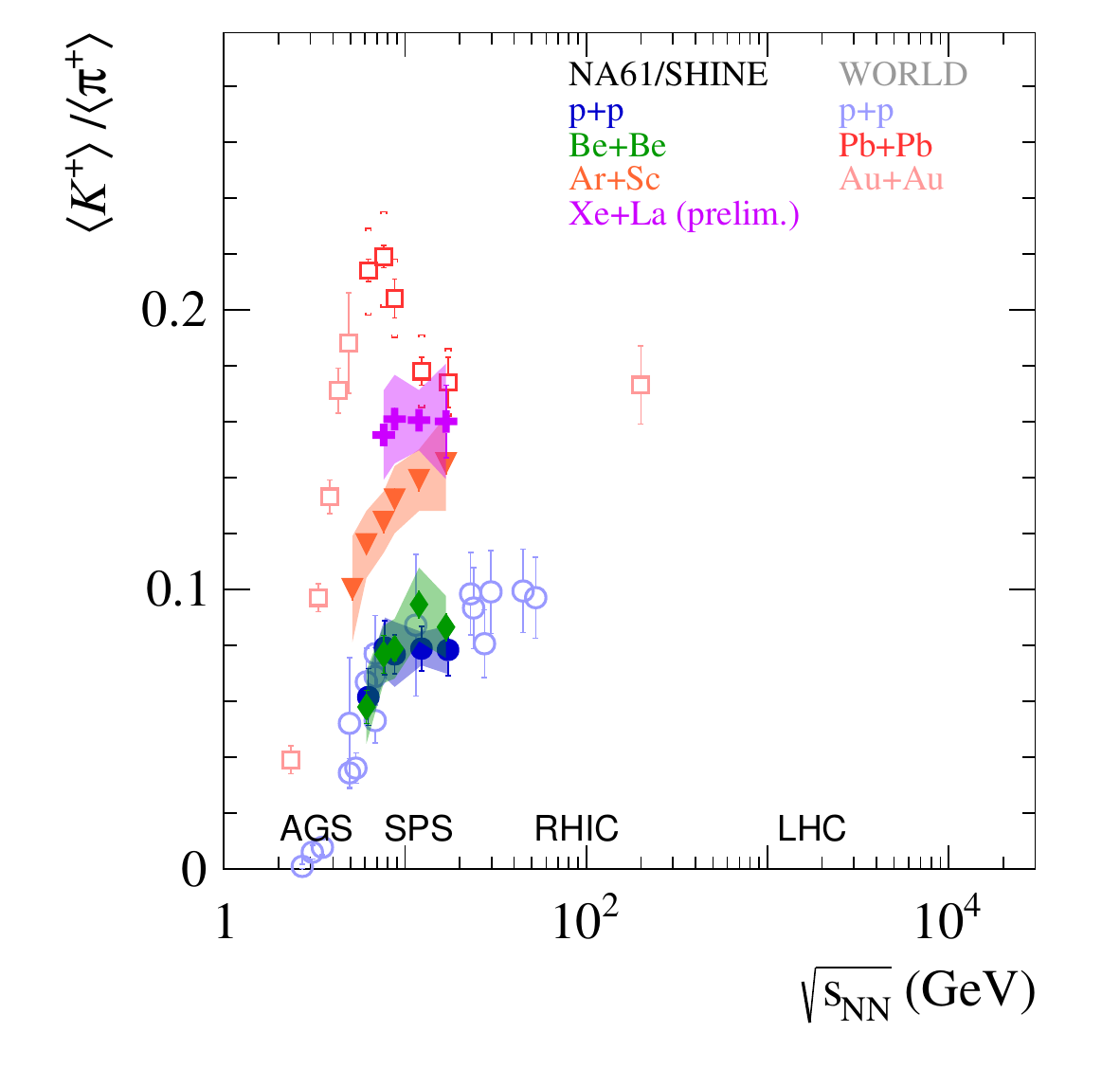}
\hspace*{-0.24cm}
\includegraphics[width=4.88cm,height=4.8cm,clip]{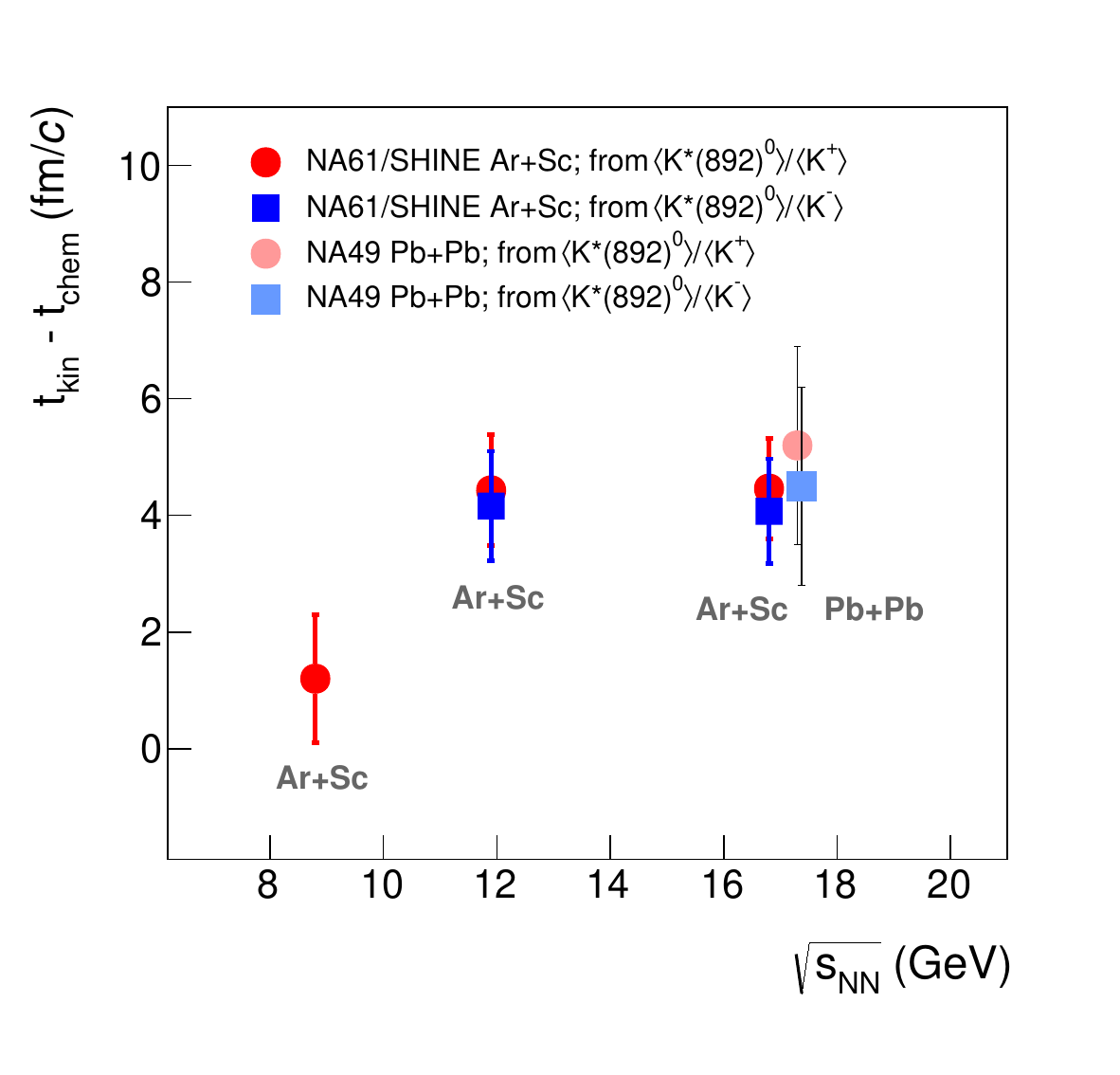}
\vspace*{-0.1cm}
\hspace*{-0.4cm}
  \begin{picture}(1,1)
 \put(-163,40){(a)}
 \put(-28,40){(b)}
  \end{picture}
  \caption{(a) Energy dependence of the $K^+/\pi^+$ ratio of 4$\pi$ mul\-ti\-pli\-cities in different systems (b) energy dependence of the time between two freeze-outs as discussed in the text.}
\label{f4}       
\vspace*{-0.2cm}
\end{figure*}
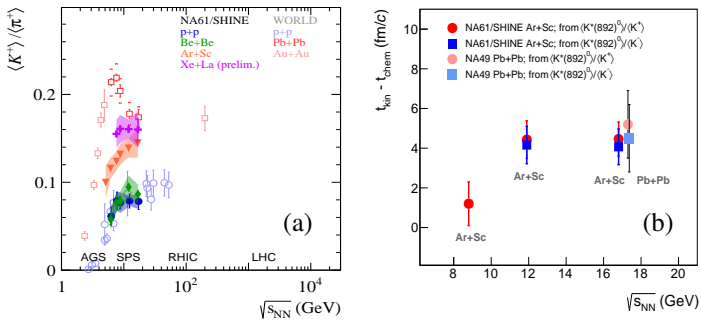

The SPS regime has a unique position in the energy and system size dependence of strangeness production. This is perfectly apparent in Fig.~\ref{f4}~(a), which shows the evolution of the $K^+/\pi^+$ ratio as a function of $\sqrt{{s}_{NN}}$ for different systems measured by NA61/SHINE with addition of other experiments, including our new preliminary data on Xe+La collisions~\cite{s}. With increasing system size, the $K^+/\pi^+$ ratio remains similar for $p$+$p$ and Be+Be reactions but rapidly enhances for Ar+Sc collisions, where it approaches Xe+La and Pb+Pb data at top SPS energy. The non-monotonic ``horn’’ in strangeness production is, at present, seen only in Pb+Pb and Au+Au reactions.

Complementary, Fig.~\ref{f4}~(b) shows the energy dependence of the estimated time between chemical and kinetic freeze-outs in Ar+Sc reactions ($t_{kin}\,$$-$$\,t_{chem}$), also compared to Pb+Pb collisions at top SPS energy. The latter estimate is evaluated on the basis of the suppression of $K^*(892)/K^{+,-}$ ratios between $p$+$p$ and central Ar+Sc (Pb+Pb) collisions, using the same methodology as in Ref.~\cite{al}. This is possible thanks to the new preliminary data on $K^*(892)$ production in Ar+Sc reactions from NA61/SHINE~\cite{bk}. A very peculiar pattern emerges: at\,$\sqrt{{s}_{NN}}$$\gtrsim$$12$~GeV the time between the two freeze-outs exceeds 4~fm/$c$, but it is consistent with zero at\,$\sqrt{{s}_{NN}}$$=$$8.8$~GeV (no suppression of the $K^*/K$ ratio in Ar+Sc compared to $p$+$p$ collisions indicates a shorter hadronic phase at this collision energy~\cite{k}).

These apparently rapid changes of system properties with system size and energy will be further quantified by NA61/SHINE, including a new energy scan of light, charge-symmetric nuclei (B+B, O+O, Mg+Mg) after LS3~\cite{10}.\footnote{For a full bibliography of NA61/SHINE and other data discussed in this section, see Refs.~\cite{3a,bk}.}
\section{Importance of non-critical effects in the search for the CP}
\label{cp}
Comprehensive theoretical reviews of the status of the search for the hypothetical critical point of the phase diagram of strongly interacting matter have been given at this Conference~\cite{10a,11}. NA61/SHINE contributes to this search by a versatile program of experimental studies bringing an altogether negative result, but at the same time demonstrating the crucial importance of a properly complete understanding of the non-critical „baseline”, that is, of spurious effects that could possibly be erroneously interpreted as
indications for
the system freezing out at the critical point.  

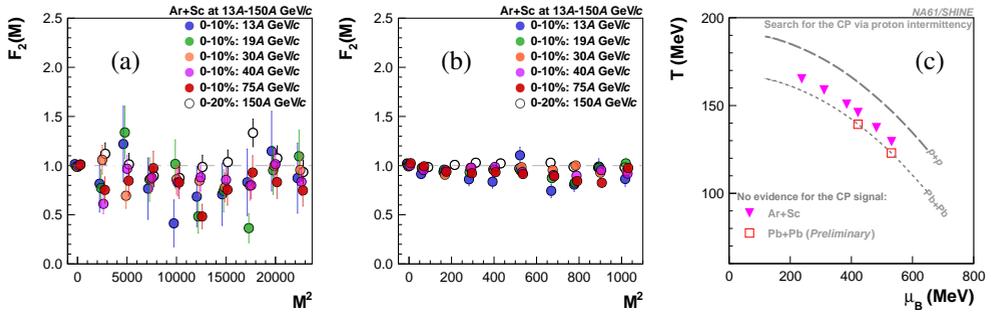
\begin{figure*}[h!]
  \centering
\hspace*{-0.7cm}
\includegraphics[width=4.89cm,clip]{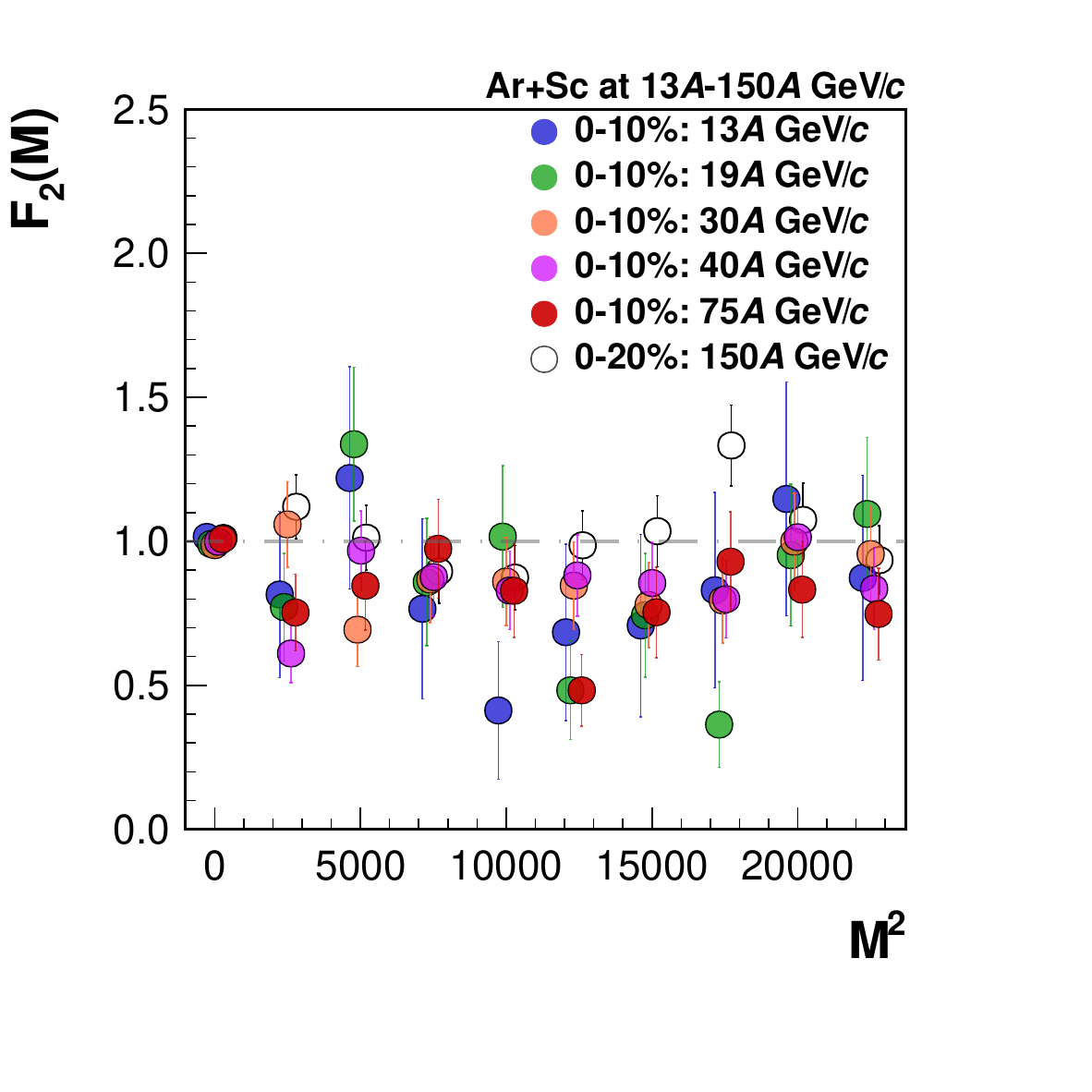}
\hspace*{-0.7cm}
\includegraphics[width=4.89cm,clip]{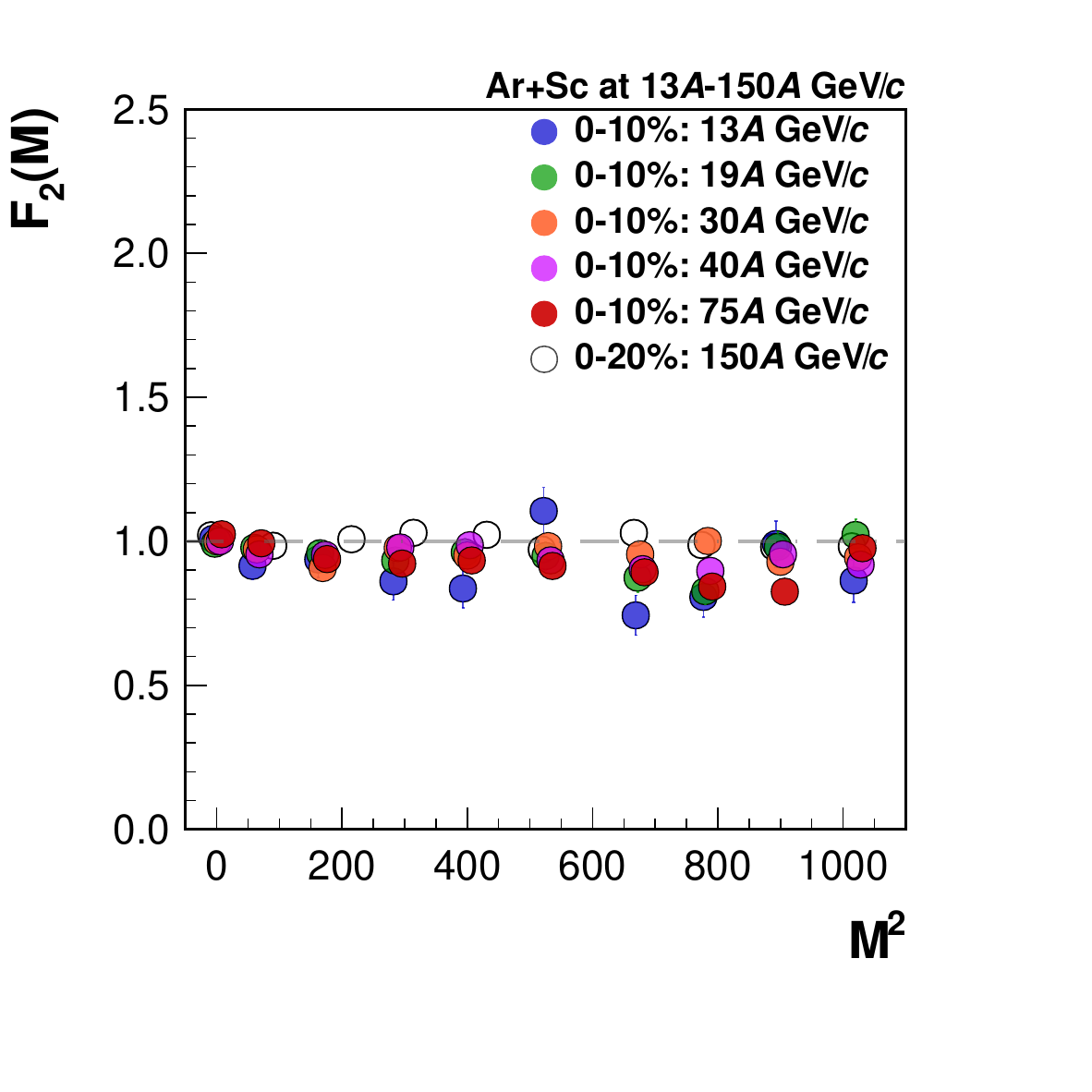}
\hspace*{-0.7cm}
\includegraphics[width=4.89cm,clip]{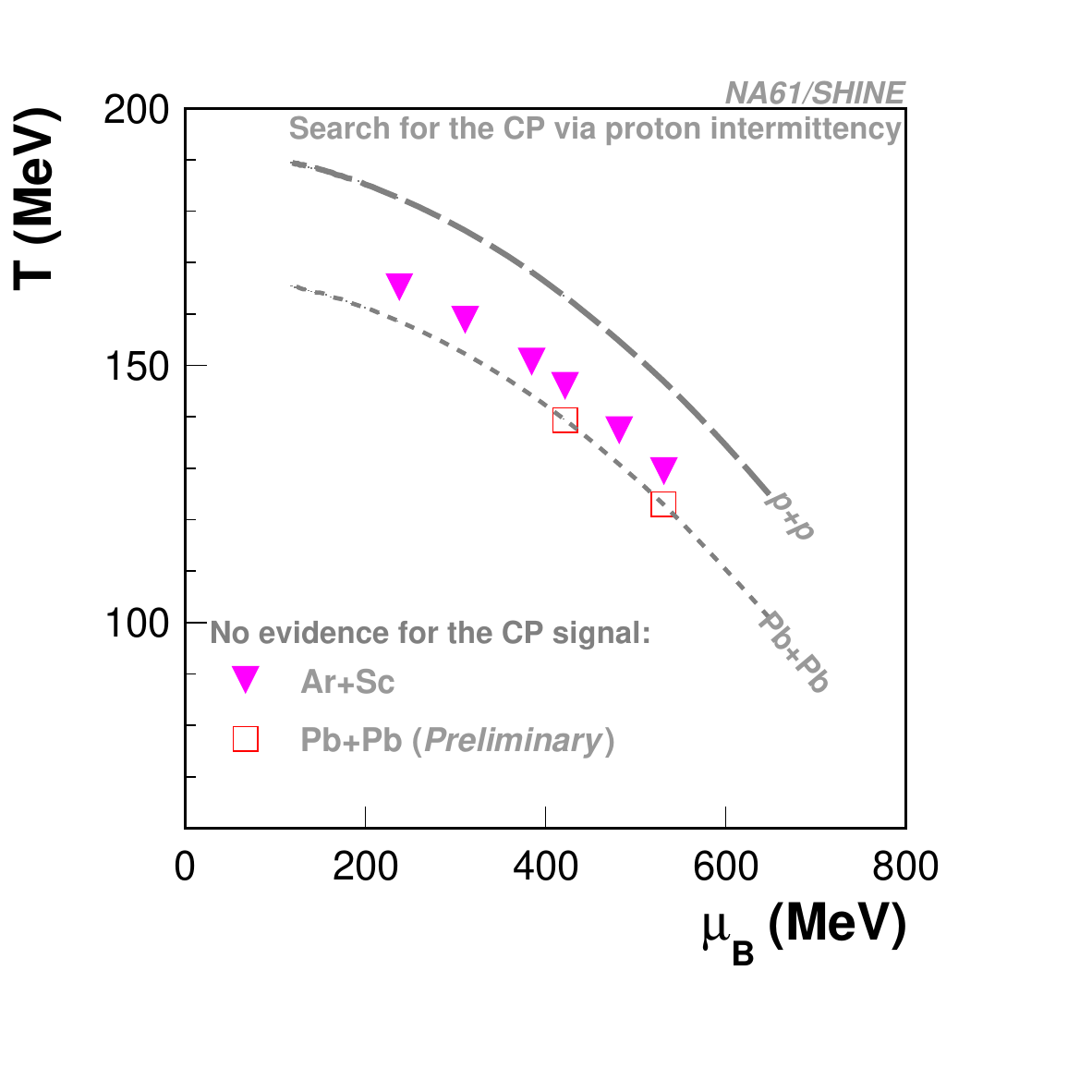}
\hspace*{-1.5cm}
\vspace*{-0.3cm}
  \begin{picture}(1,1)
 \put(-310,106){(a)}
 \put(-186,106){(b)}
 \put(-8,106){(c)}
  \end{picture}
  \caption{Proton intermittency studies in NA61/SHINE. Scaled Factorial Moments $F_2$ of proton multiplicity, drawn as a function of the number $M^2$ of cells in the cumulative transverse momentum plane for (a) $1^2\,$$\leq$$\,M^2$$\leq$$\,150^2$, and (b) $1^2\,$$\leq$$\,M^2$$\leq$$\,32^2$. Points for different energies are slightly shifted horizontally. (c) Diagram of freeze-out temperature and baryochemical potential, marking reactions where no evidence for the CP was found. All plots come from Ref.~\cite{13}. Preliminary Pb+Pb data from Ref.~\cite{15}.}
\label{f5}       
\vspace*{-0.2cm}
\end{figure*}

In Figs.~\ref{f5}~(a)-(b), new results from NA61/SHINE proton intermittency studies in Ar+Sc collisions are shown. The data comes from Refs.~\cite{13,12} where all methodological details and defi\-nitions can be found. The second-order Scaled Factorial Moments $F_2$ show no apparent power-law scaling with the number $M^2$ of cells in the transverse momentum plane, which could have been an indication of the CP. The studied reactions with no evidence for the CP cover an appreciable part of the phase diagram, see Fig.~\ref{f5}~(c). 
As described in the cited papers, this analysis employs dedicated procedures aimed at elimination of spurious, non-critical effects. This includes, among others, a cumulative transformation of protons’ ($p_x$, $p_y$) vectors which eliminates the non-scale invariant correlations but preserves the scale-invariant power-law behaviour~\cite{gb,v}.

\begin{figure*}[h!]
  \centering
\includegraphics[width=5.29cm,clip]{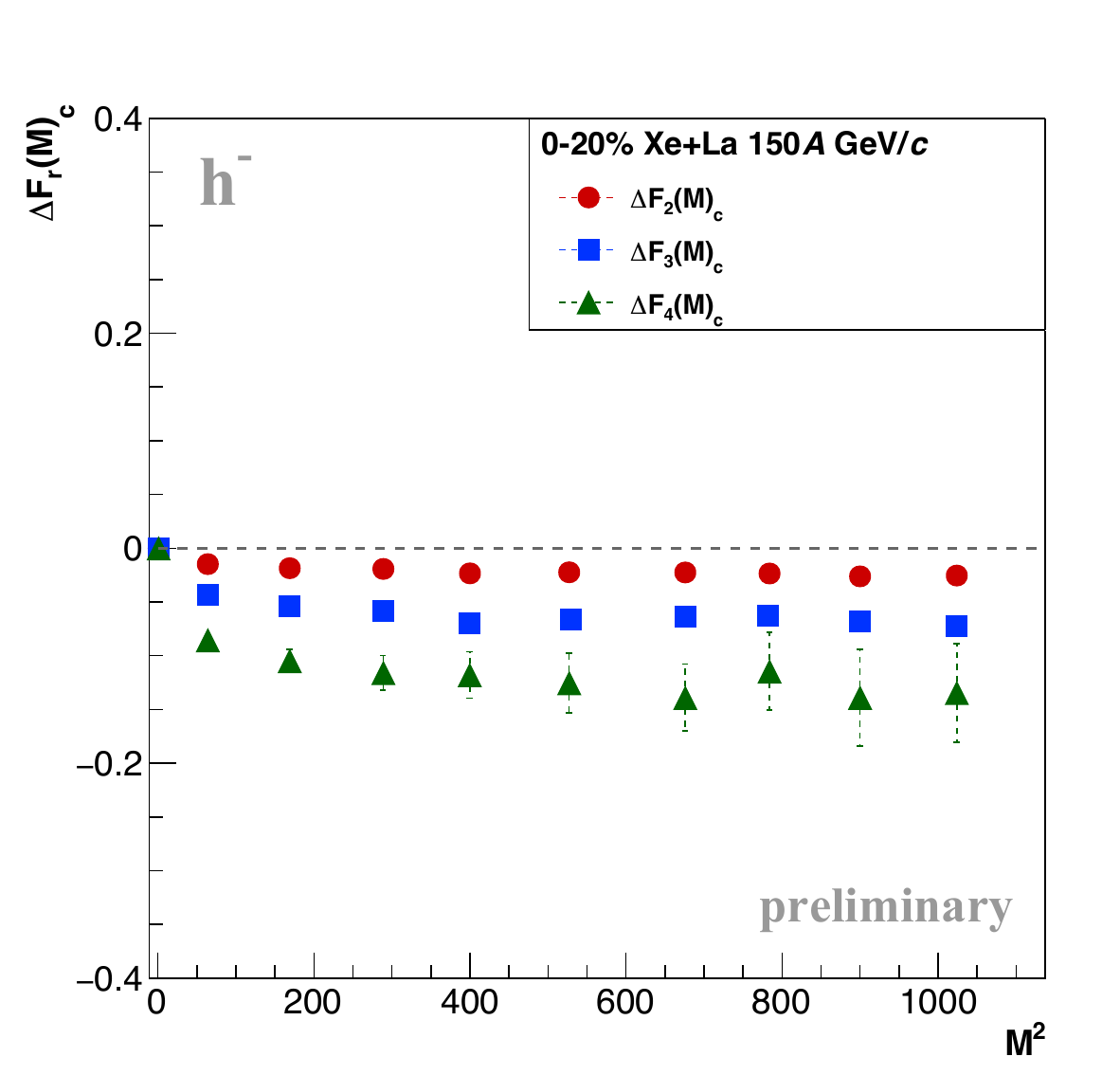}
\hspace*{-0.cm}
\includegraphics[width=5.29cm,clip]{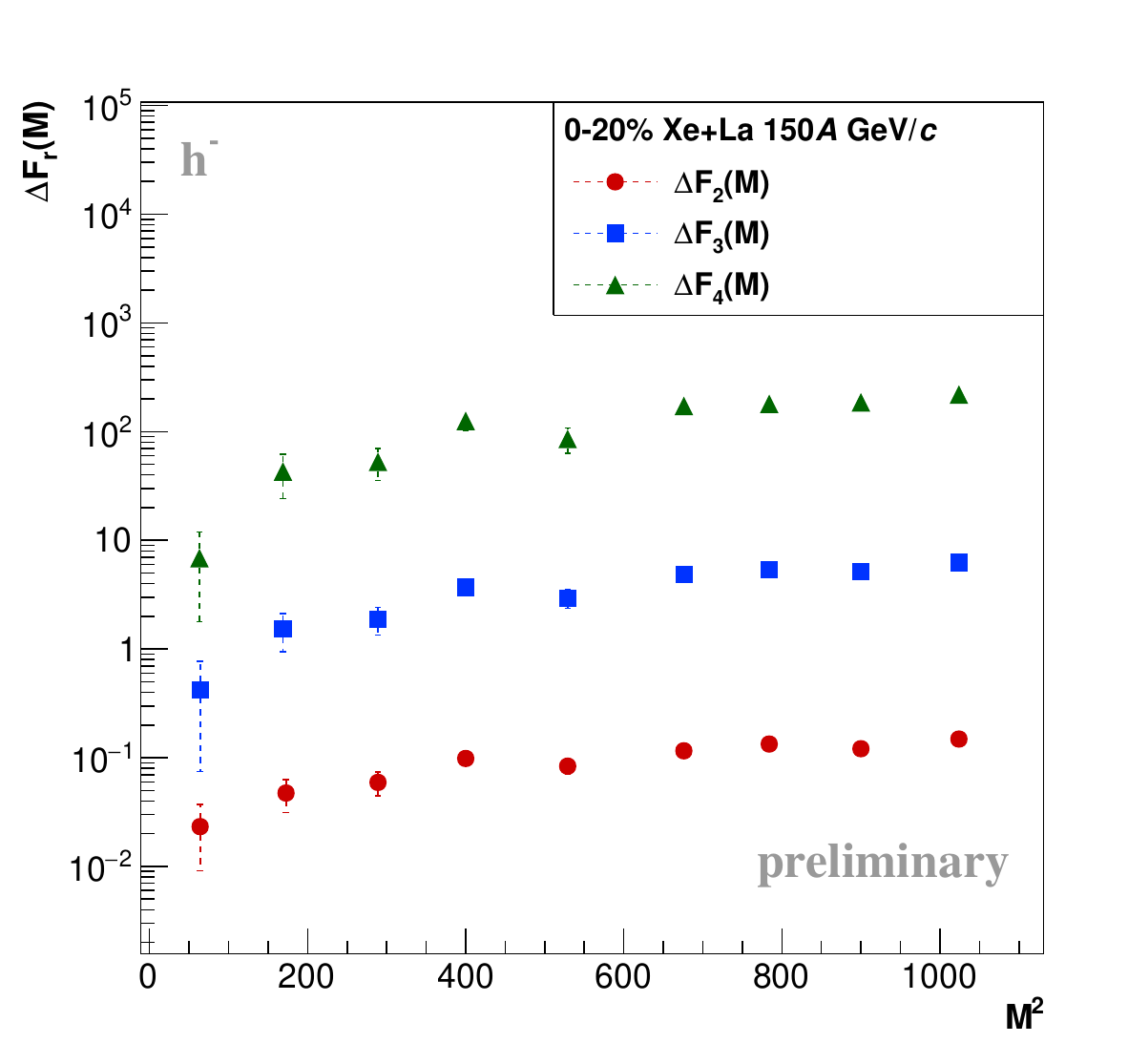}
  \begin{picture}(1,1)
 \put(-270,106){(a)}
 \put(-115,106){(b)}
  \end{picture}
  \caption{Intermittency studies of negatively charged hadrons in NA61/SHINE. The results obtained for Subtracted Scaled Factorial Moments of hadron multiplicity $\Delta F_r$ are shown up to the fourth order, for the case where the cumulative transformation of particles' ($p_x$, $p_y$) vectors (a) was applied, and (b)~was not applied to the analysis. From Ref.~\cite{v}.}
\label{f6}       
\end{figure*}
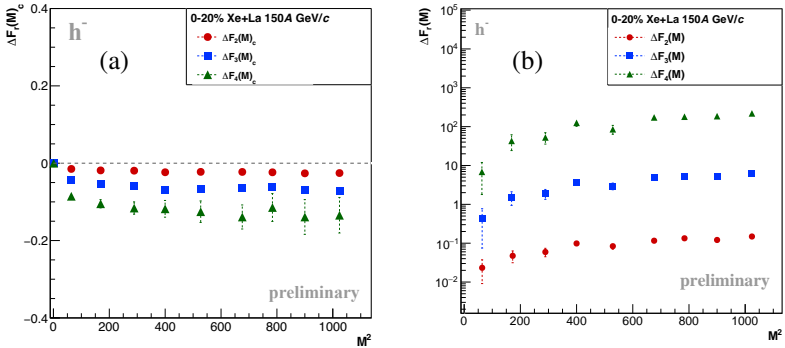

A similar study of Subtracted Scaled Factorial Moments
$\Delta F_{r}(M)_{c}$
for negatively charged hadrons produced in central Xe+La collisions is shown in Fig.~\ref{f6}\,(a), with no indication of
non-trivial
power-law scaling. For comparison, the same analysis made with {no} cumulative transformation that would eliminate spurious effects is shown in Fig.~\ref{f6}\,(b). A characteristic hierarchical structure of increase of $\Delta F_r(M)$ with $M$ and moment order $r$ emerges. This structure is qualitatively similar to that appearing in charged hadron results from STAR~\cite{plb}, also discussed at this Conference in the experimental CP review talk~\cite{z}. The fact that the dedicated procedure of cumulative transformation, Fig.~\ref{f6}\,(a), eliminates completely the characteristic structure from Fig.~\ref{f6}\,(b), has been investigated by NA61/SHINE. It has been argued~\cite{v,v1} that this structure can be explained by short-range (HBT) correlations and consequently, that the latter could also be at the source of the cited STAR result.\footnote{Note: the subtraction of uncorrelated background for $\Delta F_r$ is made differently for the cumulative and non-cumulative analyses. The non-cumulative approach relies on event-mixing, $\Delta F_r(M)=F^{data}_r(M)-F^{mix}_r(M)$~\cite{z} while for the cumulative procedure, the subtraction reduces to $\Delta F_r(M)_c=F^{data}_r(M)-F^{data}_r(1)$~\cite{v}.}

These findings clearly call for further development of experimental and theoretical methodologies for elimination of spurious (non-critical) effects in the CP search, and strongly confirm the need for a good baseline model emphasized at this Conference~\cite{11}.  For the time being, numerous models available on the market still have difficulties in description of various non-critical phenomena (see, \emph{e.g.}, section~\ref{iso} for HRG and UrQMD models). This may appear as a serious obstacle in present and future efforts to establish the presence of the critical point.
\section{Summary}
\label{sum}
The unique capabilities of the multipurpose NA61/SHINE spectrometer make it a good tool for exploitation of the versatile scientific potential of the SPS. Our most recent findings include the first-ever direct measurement of open charm production in nucleus-nucleus collisions at SPS energies, to be followed soon by
our new
series of first direct open charm measurements in a heavy ion system at the SPS. This is supplemented by the observation of an unexpected excess of charged over neutral kaon production in agreement with data from earlier experiments but with better precision. This gives evidence for violation of isospin (flavour) symmetry and calls for a concentrated experimental and theoretical effort. Findings from NA61/SHINE give no indication of the critical point, but point at the crucial importance of a proper understanding of
non-critical (baseline) effects in the CP search. \\

\small{We thank all the Organizers for their effort invested in the realization of a very successful and interesting SQM Conference (merci!). This work was supported by the Polish Ministry of Science and~Higher Education under contract No. 2021/WK/10.}

%

\end{document}